\documentclass[preprint2]{aastex}

\newcommand{\eg}{{\it e.g.,\ }}

\newcommand{\etal}{{\it et al.\ }}
\newcommand{\degs}{$^\circ$}

\newcount\notenumber
\newcount\eqnumber
\newcount\fignumber
\newbox\abstr
\newbox\figca     

\def\spose#1{\hbox to 0pt{#1\hss}}    
\def\ltsim{\mathrel{\spose{\lower 3pt\hbox{$\mathchar"218$}}
     \raise 2.0pt\hbox{$\mathchar"13C$}}}
\def\gtsim{\mathrel{\spose{\lower 3pt\hbox{$\mathchar"218$}}
     \raise 2.0pt\hbox{$\mathchar"13E$}}}

\begin{document} 
\title{\bf The Morphologies of the Small Magellanic Cloud}

\author{Dennis Zaritsky}
\affil{Steward Observatory, Univ. of Arizona, Tucson, AZ, 85721}
\affil{Electronic Mail: dzaritsky@as.arizona.edu}
\author{Jason Harris}
\affil{UCO/Lick Observatories and Board of Astronomy and
Astrophysics,} 
\affil{Univ. of California at Santa Cruz, Santa Cruz, CA, 95064}
\affil{Electronic Mail: jharris@ucolick.org}
\author{Eva K. Grebel\footnote{\noindent Hubble Fellow}}
\affil{University of Washington, Department of Astronomy, Box 351580,
Seattle, WA, 98195 }
\affil{Electronic Mail: grebel@astro.washington.edu}
\author{and}
\author{Ian B. Thompson}
\affil{Carnegie Observatories, 813 Santa Barbara St., Pasadena, CA 91101}
\affil{Electronic Mail: ian@ociw.edu}

\begin{abstract}
We compare the distribution of stars of different spectral types, and
hence mean age, within the central SMC and find that the asymmetric
structures are almost exclusively composed of young main sequence
stars. Because of the relative lack of older stars in these features,
and the extremely regular distribution of red giant and clump stars in
the SMC central body, we conclude that tides alone are not responsible
for the irregular appearance of the central SMC.  The dominant
physical mechanism in determining the current-day appearance of the
SMC must be star formation triggered by a hydrodynamic interaction
between gaseous components.  These results extend the results of
population studies (cf. Gardiner and Hatzidimitriou) inward in radius
and also confirm the suggestion of the spheroidal nature of the
central SMC based on kinematic arguments (Dopita \etal; Hardy,
Suntzeff \& Azzopardi). Finally, we find no evidence in the underlying
older stellar population for a ``bar'' or ``outer arm'', again
supporting our classification of the central SMC as a spheroidal body
with highly irregular recent star formation.
\ \ \ \ \ \ \ \ \ \ \ \ \ \ \ \ \hfill\break
\ \ \ \ \ \ \ \ \ \ 
\end{abstract} 


\ \ \ \ \ \ \ \ \ \ \ \ \ \ \ \ \hfill\break
\ \ \ \ \ \ \ \ \ \ 
\ \ \ \ \ \ \ \ \ \ \ \ \ \ \ \ \hfill\break
\ \ \ \ \ \ \ \ \ \ 
\ \ \ \ \ \ \ \ \ \ \ \ \ \ \ \ \hfill\break
\ \ \ \ \ \ \ \ \ \

\section{Introduction}

Galaxy interactions play a key role in current understanding of
galaxy formation and evolution. The dominant physical effect
of an interaction is generally thought to arise from the tidal forces
exerted between galaxies. Those tidal forces remove angular momentum
from the gas,  the gas falls toward the center of the galaxies,
and that inflow results in a nuclear, or at least centrally concentrated, 
episode of star formation (cf. Mihos, Richstone, \& Bothun 1992). 
However, for small, gaseous galaxies within
a larger dark matter halo (for which the encounter velocity is greater than
the internal velocity dispersion), 
the dominant effect of a close interaction
or collision could be hydrodynamic as the gaseous components of
the two galaxies interact. Perhaps this type of interaction is
more typical in the protogalactic environment, where small sub-galactic
stellar aggregates are coalescing into a larger galaxy, and
among interacting satellite galaxies 
within the halos of current galaxies. The nearest 
interacting satellite galaxies for study
are the Small and Large Magellanic Clouds (SMC and LMC, respectively).

The Magellanic system is highly complex and clearly interacting.
It is an ideal laboratory for examining the effects of interactions,
such as the development of tidal material and the triggering
of star formation. On the largest scales, the Magellanic Stream 
(Mathewson, Cleary, \& Murray 1974) 
is evidently a relic of an interaction, although
its origin as tidal, rather than hydrodynamic, 
is still debated because of the lack of
stars found in the stream (cf. Guhathakurta \& Reitzel 1998). 
Stellar structures in the Magellanic system that appear
to be tidal in origin were identified by Shapley (1940; the eastern SMC wing), 
by de Vaucouleurs (1955; the LMC tidal tails, although later de Vaucouleurs
(de Vaucouleurs \& Freeman 1972) partially retracted his claim due to potential confusion with
galactic diffuse emission), by  Hindman, Kerr, \& McGee (1963; the H I 
bridge between the
LMC and SMC), by Irwin, Demers, \& Kunkel (1990; the stellar bridge
between the LMC and SMC), and by Putman \etal (1998; the leading Magellanic
Stream).
Recent numerical simulations (only including gravity; Gardiner and 
Noguchi 1996) show that such dynamics reproduce 
many of the observed features, including the SMC wing and the 
line-of-sight depth of the outer regions of the SMC (Hatzidimitriou and 
Hawkins 1989). 

Our study of the distribution of stars in the SMC
is complementary to that of the outer SMC by 
Gardiner and Hatzidimitriou (1992). The structure of the central
SMC is currently somewhat more uncertain than that of the outer 
regions (cf. Hatzidimitriou, Cannon, and
Hawkins 1993 for a discussion). We demonstrate, 
based on the distribution of stars of different ages
within the central SMC (4\degs $\times$ 4\degs), that although the existence
of tidal forces on the SMC is not in dispute, the visible appearance of the
central SMC arises primarily from hydrodynamic effects.
The underlying, old stellar population of the central SMC
is relatively undisturbed (as concluded by Dopita \etal 1985; Hardy \etal 1989 
on the basis of kinematic arguments). These
results provide direct evidence of star formation triggered by the interaction
of the SMC with another gaseous system (presumably, but not demonstrated
to be, the LMC). Several common features, such as the ``bar''
and the ``outer arm'' are not seen in the underlying population and
so are not true dynamical structures.

\section{The Data}

The data come from the ongoing Magellanic Cloud Photometric
Survey (cf. Zaritsky, Harris, \& Thompson 1997). Using the
Las Campanas Swope telescope (1m) and the
Great Circle Camera (Zaritsky, Shectman, \& Bredthauer 1996)
we have been drift scanning both Magellanic Clouds in $U, B, 
V, $ and $I$. The effective exposure time is between 4 and 5 min for SMC
scans and the pixel scale is 0.7 arcsec/pixel. 
The data are reduced using a pipeline that utilizes
DAOPHOT (Stetson 1987) and
IRAF\footnote{IRAF is distributed by the National Optical Observatories,
which are operated by AURA Inc., under contract to the NSF}. Only stars with both $B$ and
$V$ detections are included in the catalog. A complete description
of the SMC data will be presented by Zaritsky \etal (2000).

A V-band luminosity map of the SMC is constructed from the photometric
catalog of stars
with $m_V < 20$ and shown in Figure 1. 
Several well-known, distinctive features
are evident. First, there is a general elongation from North-East
to South-West, most noticeably in the central region that is often
referred to as the ``bar'' of the SMC. Second, there are high surface
brightness knots  directly to the east of the main body of the SMC that
are also accompanied by a distortion of the fainter SMC isophotes.
These features are part of the inner SMC
wing, first identified by Shapley (1940), that extends
to the bridge and intercloud region between the SMC and LMC (Irwin,
\etal 1990). 
The general impression from this Figure is that the SMC is 
an irregular system, possibly with a bar, and little, if any,
spiral structure.

By having a stellar catalog rather than the images alone,
we can separate different stellar populations and independently examine their
distributions. First, we select upper main
sequence stars ($m_V < 18.5, M_V < -0.4$ , for $m-M = $ 18.9
and $-0.3 < B-V < 0.3$, which corresponds to an age $\ltsim 
2 \times 10^8$ years)
and show their spatial density 
distribution in Figure 2. The greyscale value of each pixel corresponds
to the number of stars within that pixel, with dark pixels indicating
a high density of stars. Here we see
a different morphology than that illustrated in Figure 1. 
The NE-SW elongation is more marked, there
is a pronounced extension of these stars toward the extreme SW that
was not visible in Figure 1, a concentration in the 
NE that appears to be a distinct linear feature perpendicular
to the main elongation of the SMC, 
and the knots in the wing are relatively more prominent than in Figure 1.
The young stars extend along three different directions toward
the edges of the survey regions and apparently beyond the survey 
region toward the East along the wing (cf. Grondin, Demers,
\& Kunkel 1992; Demers \& Battinelli 1998) and toward the SW. The system
of young stars is highly disturbed and somewhat concentrated toward the
body of the SMC, but it extends to large angular distances. The young
star distribution is qualitatively
similar to the H I distribution (Stanimirovic \etal 1999).

In contrast, our second selected stellar population (giants and red
clump stars, $m_V < 19.5, M_V < 0.6$ and $B-V > 0.7$, or correspondingly
stars with ages $\gtsim 1$ Gyr) is highly
regular (Figure 3).  There is no evidence of the wing, the SW
extension or the NE shell. There are distinct differences in stellar
densities among scans, mostly due to variable seeing conditions and
hence catalog completeness levels. We have reobserved regions with poor
($>$ 2 arcsec) seeing, but those data are not included here. We have
also obtained data to fill in the gaps between scans, but again those 
are not yet available. A complete discussion of the surface
brightness distribution of the SMC with the final data 
will be given elsewhere,
but it is evident that the distribution of the older stellar population is
much more regular than that of the younger population. 
Tidal forces have not noticeably
distorted the projected morphology of the older, central SMC
population. The only potential distortion of the older
population appears to be a subtle asymmetry
toward the South, but we cannot yet determine if this is an artifact of the 
scan-to-scan variations. The conclusion to be derived from Figure 3 
is that the
features that one might have thought were tidally induced (the wing
and the NE and SW extensions) are not present in the distribution
of old ($>$1 Gyr) stars, and so cannot solely originate from tidal effects
(i.e.  normal stellar
populations extracted from the central SMC by tidal forces).

\section{Discussion}

There are several aspects of the observations that appear
in conflict with the hypothesis of 
a tidal origin for the morphology of the
system. First, the recent star formation is more extended than the 
older stellar system, rather than being more concentrated as
most generic interaction models suggest. Second, the older stars do not
follow what initially appear to be the tidal ``tails'' as noticeably
as the younger stars. There are old stars at large radius from the
SMC, but their distribution is much 
more regular than that of the younger stars (Gardiner and Hatzidimitriou 
1992). One
must conclude either that the tides predominantly affected the outer
part of the SMC, which presumably contained large amounts
of gas, that then formed the young stars --- or that star formation
in the central SMC was triggered when the SMC
interacted directly with a gaseous component (either the LMC gaseous
halo or perhaps a third body). 

The outer tidal hypothesis has several difficulties. First, 
the stellar extensions appear to go in several directions rather
than the standard tail/bridge geometry. Second, if the gaseous
material was originally at large radius and of insufficient density to form
stars, one would expect that tidal 
forces would further lower the densities at large radii where
the gas has no means to lose angular momentum and dissipate
energy. In its favor, at least
as an explanation for the wing structure, is the continuity between the
wing and bridge (both in stars and gas) that join the SMC and LMC.
But if the dominant tidal force created the LMC-SMC bridge, what 
formed the NE-SW structure of young stars in the SMC?

We speculate that the NE-SW geometry
arises either from shocking of the gas in the central SMC as the SMC
moves in a perpendicular direction (i.e. NW or SE) to a second gaseous
object (\eg speculatively, 
a hot Milky Way halo or the outer gaseous envelope of the 
LMC), or that the infall of a gaseous cloud along the NE-SW axis
formed stars that have orbits aligned with the infall axis.
We see no evidence for shocking in the H I map of Stanimirovic
\etal 1999, although the H I morphology is highly irregular.
Furthermore, the H I kinematics contain a large velocity gradient
along the ``bar" axis, which would not arise naturally in a 
ram pressure model but which could arise in a model of the collision of two
gaseous components along the ``bar" axis.
The latter hypothesis
is also supported by the presence of the stellar ``shell" feature (upper
left, Figure 2), which could arise 
if young stars are on radial orbits along the major
axis of the distribution. This discussion is highly speculative and
a definite conclusion awaits numerical simulations that include both
gravity and hydrodynamics, and additional measurements of the stellar kinematics.

Despite the disturbed visual appearance of the SMC and the observations
supporting large line-of-sight depths in the outer SMC, the bulk of the stars 
in the SMC's central region evidently form a spheroidal population. 
This result confirms previous observations of the kinematics of older stars.
Dopita \etal 1985 from PN observations and Hardy \etal 1989 from C star 
observations found that the older stellar component in the central SMC region 
has the kinematics of a spheroidal component, with no clear signs for
multiple kinematic components or strong rotation. In contrast, the
H I shows a large kinematic gradient along the NE-SW axis (Stanimirovic
\etal 1999). Furthermore, we find no evidence for a bar in  the underlying 
older stellar 
distribution and the ``outer arm'' is entirely a product of the young 
main sequence stars. 

The discrepancies between morphological classifications
based on images, which inappropriately weigh the younger populations,
and stellar catalogs, from the which the numerically dominant,
underlying population
can be extracted, highlight the inherent difficulties in
morphological studies from integrated photometry. 
These difficulties are exacerbated in
studies at higher redshifts, where the younger stars are 
even more disappropriately weighted due to K-corrections and where the
spatial resolution is poorer. 

We conclude, 1) that the central SMC is principally a spheroidal system,
2) that its visual morphology is dominated by highly irregular recent
star formation, and 3) that hydrodynamics, rather than tidal
forces, must play the key role in  SMC star formation over the last
several hundred million years.

\vskip 1in
\noindent
ACKNOWLEDGMENTS: DZ acknowledges partial financial support from an
NSF grant (AST-9619576), a NASA 
LTSA grant (NAG-5-3501), and fellowships from
the David and Lucile Packard Foundation and the Alfred P. Sloan
Foundation. EKG acknowledges support from NASA through 
grant HF-01108.01-98A from the Space Telescope Science Institute.
\vskip 1cm
\noindent
\vfill\eject

\onecolumn

\begin{figure}
\includegraphics{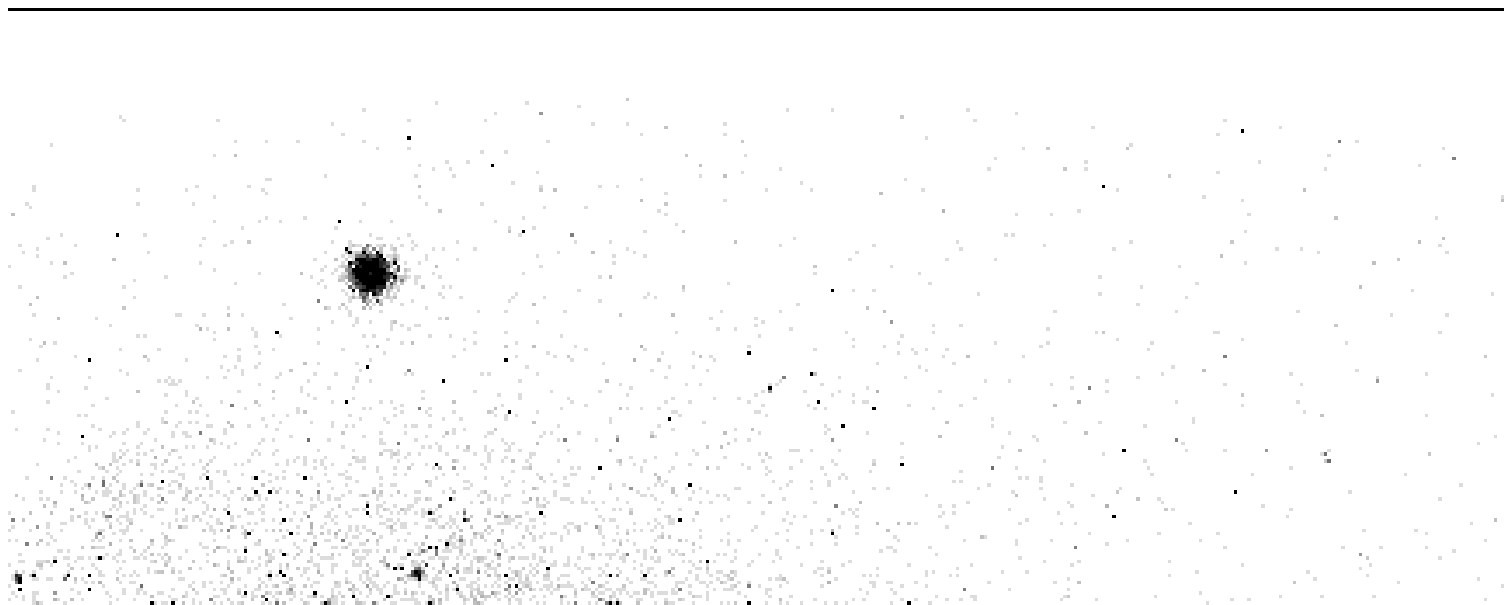}
\vskip 6in
\caption{The stellar luminosity map of the SMC recreated from stars in
the catalog with $M_V < 20$. The image is $4^\circ \times
4^\circ$. The outer halo of 47 Tuc is just visible off the right side
of the image and NGC 362, a Galactic foreground cluster, is in the
upper left. The orientation is standard, with North up and East
left. The isolated foreground stars, have been removed using a local
median algorithm.  The brighter objects remaining that appear to be
stars are most likely SMC clusters.}
\end{figure}
\clearpage
\begin{figure}
\includegraphics{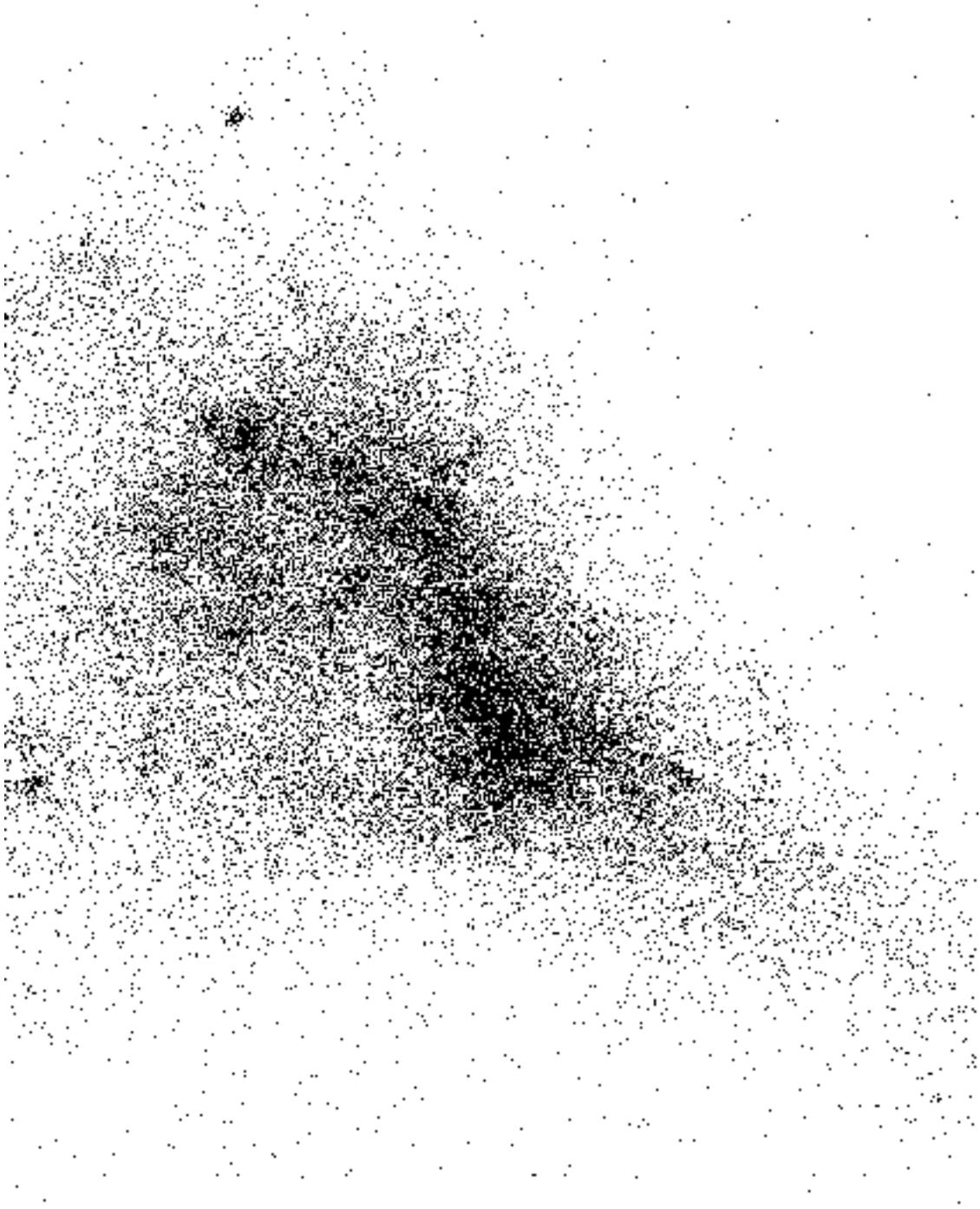}
\vskip 6in
\caption{The stellar density map of the SMC using only the upper
main sequence stellar population ($V < 18.5$ and $-0.3 < B-V < 0.3$,
which corresponds to age $\ltsim 2 \times 10^8$ yrs). The
angular scale and orientation are the
same as in Figure 1.  112,107 stars contribute to this Figure.}
\end{figure}
\clearpage
\begin{figure}
\includegraphics{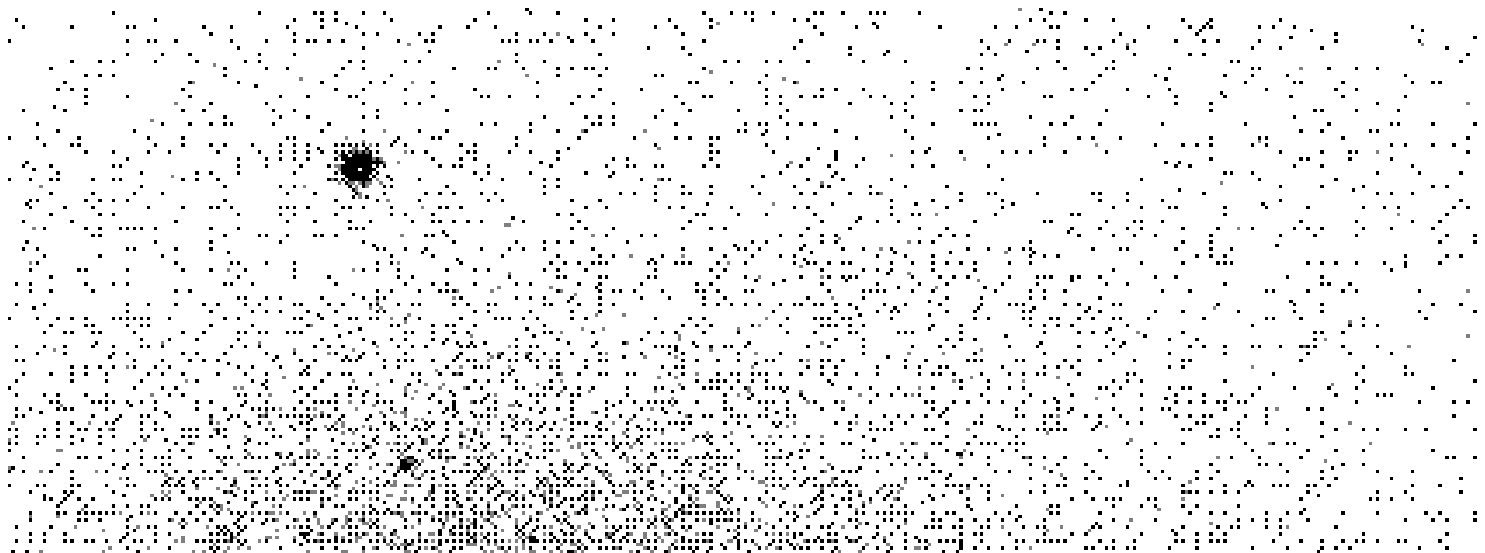}
\vskip 6in
\caption{The stellar luminosity map of the SMC using only the 
red giant and red clump stellar populations (age $\gtsim 1$ Gyr). 
The spatial scale and 
orientation are the same as in Figure 1.  686,742 stars contribute
to this Figure.}
\end{figure}
\clearpage

\begin{thebibliography}{}
\bibitem[d55]{d55}de Vaucouleurs, G. 1955, \aj, 60, 126
\bibitem[d72]{d72}de Vaucouleurs, G. and Freeman, K.C. 1972, Vistas in Astronomy, 14, 163
\bibitem[d98]{d98}Demers, S. \& Battinelli, P. 1998, \apj, 115, 154
\bibitem[d85]{d85}Dopita, M.A., Ford, H.C., Lawrence, C.J., and 
   Webster, B.L. 1985, \apj, 296, 390
\bibitem[k76]{k76}Kunkel, W.E., \& Demers, S. 1976, Roy. Green. Obs. Bull.,
   182, 241
\bibitem[g92]{g92}Gardiner, L.T., \& Hatzidimitriou, D. 1992 \mnras, 257, 195
\bibitem[g96]{g96}Gardiner, L.T., \& Noguchi, M 1996, \mnras, 307, 577
\bibitem[g92]{g92}Grondin, L., Demers, S., \& Kunkel, W.E. 1992, \aj, 103, 1234
\bibitem[g98]{g98}Guhathakurta, P., \& Reitzel, D.B. 1998, in Galactic Halos:
   A UC Santa Cruz Workshop, ed. D. Zaritsky (Astronomical Society of
   the Pacific: San Francisco), p. 22
\bibitem[h89]{h89}Hardy, R., Suntzeff, N.B., \& Azzopardi, M. 1989, \apj, 344,
   210
\bibitem[h93]{h93}Hatzidimitriou, D., Cannon, R.D., and Hawkins, M.R.S. 1993, 
   \mnras, 261, 873
\bibitem[h89]{h89}Hatzidimitriou, D., \& Hawkins, M.R.S. 1989, \mnras 241, 667
\bibitem[h63]{h63}Hindman, J.V., Kerr, F.J., \& McGee, R.X. 1963, Aust. J. 
   Phys., 16, 570
\bibitem[i90]{i90}Irwin, M.J., Demers, S., \& Kunkel, W.E. 1990, \aj, 99, 191
\bibitem[m74]{m74}Mathewson, D.S., Cleary, M.N., and Murray, J.D. 1974, \apj, 
   190, 291
\bibitem[m92]{m92}Mihos, J.C., Richstone, D.O., \& Bothun, G.D. 1992, \apj,
   400, 153
\bibitem[p98]{p98}Putman, M.E., \etal 1998, Nature, 394, 752
\bibitem[s40]{s40}Shapley, H. 1940, Harvard Obs. Bull. 814, 8
\bibitem[s99]{s99}Stanimirovic, S., Staveley-Smith, L., Dickey, J.M.,
Sault, R.J., \& Snowden, S.L. 1999, \mnras, 302, 417
\bibitem[s87]{s87}Stetson, P.B. 1987, \pasp, 99, 191
\bibitem[z97]{z97}Zaritsky, D., Harris, J., \& Thompson, I 1997, \aj, 
   114, 1002
\bibitem[z00]{z00}Zaritsky, D., Harris, J., Grebel, E.K., \& Thompson, I 2000, \aj, in prep.
\bibitem[z96]{z96}Zaritsky, D., Shectman, S.A., \& Bredthauer, G 1996, \pasp,
   108, 104
\end{thebibliography}
\end{document}